\documentclass[]{aa}

\usepackage{graphicx}
\usepackage{txfonts}

\newcommand{\vsini}{$\varv\sin i$}

\newcommand{\vp}{$\varv_{\rm{p}}$}
\newcommand{\vc}{$\varv_{\rm{c}}$}

\newcommand{\sbs}{SB1$^{*}$}
\newcommand{\logl}{$\log L$}
\newcommand{\logfc}{$\log f_{\rm c}$}
\newcommand{\logf}{$\log f$}
\newcommand{\logfp}{$\log f_{\rm p}$}
\newcommand{\fpe}{$f_{\rm p}$}
\newcommand{\fci}{$f_{\rm c}$}
\newcommand{\fl}{$f_{\rm L}$}
\newcommand{\rl}{$R_{\rm L}$}

\newcommand{\kms}{{km\,s}$^{-1}$}

\newcommand{\teff}{$T_\mathrm{eff}$\,}

\newcommand{\Msun}{$M_\odot$}
\newcommand{\Lsun}{$L_\odot$}
\newcommand{\Rsun}{$R_\odot$}
\newcommand{\Tsun}{$T_\odot$}

\begin{document}

   \title{Rotational synchronisation of B-type binaries in 30 Doradus}
   \titlerunning{B-type binary synchronisation}


   \author{D.~J. Lennon\inst{1,2},
        P.~L. Dufton\inst{3},
        J.~I. Villase\~nor\inst{4},
        N. Langer\inst{5},
        C.~J. Evans\inst{6},
        H. Sana\inst{7},
        W. D. Taylor\inst{8}
          }
\authorrunning{D.~J.~Lennon et al.} 

   \institute{ 
         Instituto de Astrof\'isica de Canarias, E-38\,200 La Laguna, Tenerife, Spain
         \and
         Dpto. Astrof\'isica, Universidad de La Laguna, E-38\,205 La Laguna, Tenerife, Spain
         \and
         Astrophysics Research Centre, School of Mathematics \& Physics,  Queen’s University, Belfast, BT7 1NN, UK 
         \and
         Max-Planck-Institut f\"ur Astronomie, K\"onigstuhl 17, D-69117 Heidelberg, Germany         
         \and
         Argelander-Institut f\"{u}r Astronomie der Universit\"{a}t Bonn, Auf dem H\"{u}gel 71, 53121 Bonn, Germany
         \and
         European Space Agency, ESA Office, Space Telescope Science Institute, 3700 San Martin Drive, Baltimore, MD 21218, USA
         \and
         Institut voor Sterrenkunde, Universiteit Leuven, Celestijnenlaan 200 D, B-3001 Leuven, Belgium
         \and
         UK Astronomy Technology Centre, Royal Observatory Edinburgh, Blackford Hill, Edinburgh, EH9 3HJ, UK
    }

   \date{}

 
  \abstract
  {The spin evolution of stars in close binary systems can be strongly affected by tides. We investigate the rotational synchronisation of the stellar components for 69 SB1 systems and 14 SB2 B-type systems in the 30 Doradus region of the Large Magellanic Cloud using observations from the VFTS and BBC surveys. 
  Their orbital periods range from a few to a few hundred days, while estimated primary masses for these systems are in the range $\sim$5--20\,\Msun\ with mass ratio ranges of $q\sim$0.03--0.5 and $q\sim$0.6–-1.0 for the SB1 and SB2 systems, respectively. Projected rotational velocities of the stellar components have been compared with their synchronous velocities derived from the orbital periods. We find that effectively all systems with orbital period of more than 10 days must be asynchronous, whilst all the systems with periods of less than 3 days are likely synchronised. In terms of the stellar fractional radius (r), our results imply that all systems with r < 0.1 are asynchronous, with those having r > 0.2 probably being synchronised. For the apparently synchronised systems our results are more consistent with synchronisation at the mean orbital angular velocity rather than with that at periastron.
  }


   \maketitle



\section{Introduction}\label{s_intro}
Most early-type stars are believed to reside in multiple systems \citep{mas09}. For example, \citet{san12}
estimated an intrinsic binary fraction of 69\% for Galactic O-type systems, with a strong preference for closely bound systems. Other Galactic studies \citep[see, for example, ][]{mah09, mah13, kob14, pfu14, ald15, gra23} have found similar fractions for early-type stars. In the Large Magellanic Cloud (LMC),  \citet{san13} estimated a fraction of 51$\pm$4\% for the O-type stellar population of the 30~Doradus region observed in the VLT-FLAMES Tarantula survey \citep[VFTS][]{eva11}, whilst \citet{dun15} inferred a similar high fraction, 58$\pm$11\%, for the corresponding B-type stellar population. Both of these binary sub-samples were followed up with multi-epoch campaigns aimed at characterising their orbits in the Tarantula Massive Binary Monitoring programme \citep[TMBM;][]{alm17} for the O-type primaries, and the B-type Binary Characterisation programme \citep[BBC;][]{vil21} for systems with B-type primaries.

\citet{deM14} undertook theoretical simulations that implied that some massive stars (typically $\sim$8\%) might be the products of stellar mergers. In turn this would imply an even higher fraction of early-type stars have been formed in binary systems. Recently \citet{men24} concluded that a large fraction of the LMC single blue supergiant population were also the products of binary mergers.

The evolution of these binary systems will depend both on their initial parameters and how these evolve with time. Here we investigate the rotational velocities of B-type binary components in the LMC, and in particular the degree of synchronisation of their rotational and orbital periods. This will influence how the stellar components evolve, including the amount of rotational mixing that they undergo. For example, \citet{duf23} found that the relatively small nitrogen abundance estimates for  Be-type stellar atmospheres \citep{len91, dun11, duf23} were consistent with their creation via a binary evolutionary channel.

Although there are studies of the synchronisation of individual massive systems \citep[see, for example,][]{lan17, put18, and19, put21}, there are relatively few large scale studies. \citet{lev76} considered 122 system with primary spectral type earlier than G0. For the B-type sub-sample of 26 systems, those with periods less than 4 days appeared to be synchronised but there was evidence that this limit increased with the increasing stellar age. \cite{giu84ecl} studied 140 eclipsing binaries that ranged in spectral type from effectively O7 to F6 (with one component classified as K-type). Defining the stellar fractional radius $r$ as the ratio of the stellar radius ($R$) to the system semi-major axis ($a$), i.e. $r=R/a$, they found that almost all members were synchronised down to a fractional radius of $r \sim 0.15$ with both synchronised and asynchronous rotators being found down to a fractional $r \la 0.10$. There were insufficient systems to discuss the degree of synchronisation for $r < 0.1$. 

Subsequently \citet{giu84e, giu84l} considered early- and late-type non-eclipsing systems. For the early-type system (with spectral types from O8 to A4), \citet{giu84e} found that 'deviations from synchronous rotation were most apparent at fractional radii of 0.05 and below'. Recently \citet{zim17} and \citet{sai22} have used the r-mode oscillations in approximately 20 Kepler binary systems to deduce the stellar rotational periods of the primary (with estimated masses between 1.2 and 2.3\Msun). Their conclusions on the degree of synchronisation differed and this is further discussed in Sect.\ \ref{d_pre}

 In this paper, we discuss the binary systems with B-type primaries that have been identified in VFTS \citep[][]{eva11,eva15, dun15} and orbital parameters derived from BBC \citep[][]{vil21}. Estimated orbital periods have been combined with estimates of the projected rotational velocities \citep{gar17} to investigate whether the stellar components could be synchronised. We then discuss our results in terms of both the orbital periods and stellar separations and also compare with theoretical predictions.

\section{Observational sample}

\citet{dun15} searched for multiplicity amongst more than 400 B-type stars observed in the VFTS survey \citep{eva11,eva15} and identified 101 targets as possible SB1 or SB2 systems on the basis of significant variability in their radial velocity estimates or doubling of their spectral features.  \citet{vil21} obtained additional spectroscopy for eighty eight of these targets and characterised them as:
\begin{enumerate}
    \item SB1: definitely single line spectroscopic binary
    \item \sbs: possibly single line spectroscopic binary
    \item SB2: definitely double lined spectroscopic binary
    \item RV variable: radial velocity variable with status unclear
\end{enumerate}
We have considered all sources in the SB1, SB1* and SB2 groups except for two Be-type systems in the \sbs\ group: VFTS\,697 and 847. These two apparently short period binaries were excluded as attempts to improve their orbital solutions by the addition of VFTS epoch data to the BBC data failed to confirm the already uncertain results,  As discussed in \cite{vil21} the low-level radial velocity variability is likely related to intrinsic variability of the Be-type stars. 
The final sample therefore contains 51 SB1, 18 \sbs\ and 14 SB2 systems\footnote{an additional B-type SB1 system (\#829) from \citet{alm17} has been included}, that constitute the largest sample with B-type primaries to be analysed and their properties are summarised in Tables \ref {t_SB1_full} and \ref{t_f_sb2} with the photometry and spectral types taken from \citet{eva11}.

\section{SB1/\sbs\ sample}

\subsection{Projected rotational velocities, \vsini} \label{s_vsini_sb1}

Projected rotational velocity (\vsini) estimates for the targets characterised as SB1/\sbs\ were taken mainly from \citet{gar17}. These were estimated using a Fourier Transform (FT) technique \citep{car33, sim07} following the procedures described in \citet{duf12}. This approach has been widely used for early-type stars \citep[see, for example][]{dufsmc06, lef07, mar07, sim10, fra10, duf12, sim14, sim17} and relies on the convolution theorem \citep{gra05}, viz. that the Fourier transform of convolved functions is proportional to the product of their individual Fourier Transforms. It then identifies the first minimum in the Fourier transform for a spectral line, which is assumed to be the first zero in the Fourier transform of the rotational broadening profile with the other broadening mechanisms exhibiting either no minima or only minima at higher frequencies. Further details on the implementation of this methodology are given by \citet{sim07, duf12, gar17}. Projected rotational velocities were also obtained from fitting observed lines with rotational broadening functions \citep{gra05}. These estimates should be considered as upper limits as they did not include other broadening mechanisms. However as discussed by \citet{duf12} and \citet{gar17}, they were consistent with the estimates obtained by the FT methodology and indeed the principle results discussed below would have remained unchanged if we had adopted these values.

For seven supergiant systems (\#027, 430, 576, 687, 779, 827, 829) the  \vsini-estimates were taken from \citet{mce15}, who used the {\sc iacob-broad} tool as discussed in \citet{sim14}. This tool performs a complete line-broadening analysis  yielding \vsini-estimates via a FT methodology and macroturbulent broadening via a line-profile fitting techniques. The latter may be significant in these supergiant spectra but the theoretical profiles containing both macroturbulent and rotational broadening gave convincing fits to the observations.

\citet{gar17} provided standard deviations for the \vsini-estimates of each star, which can be used to estimate the standard error of the means assuming a normal distribution. These have a median of 4.6\,\kms\ and range from 1.5 to 9.2\,\kms\ depending on the quality of the spectroscopy and the number of lines measured. The corresponding percentage uncertainties have a median of 4.4\% and range from 0.9 to 13.1\%. For the supergiant targets \citet{mce15} recommended conservative error estimates of 10\% for \vsini$\geq$100\,\kms and 20\% for \vsini$<$100\,\kms, due to the significant contribution of macroturbulence to the line broadening for the latter

\subsection{Synchronous velocities}\label{s_f_sb1}

Synchronous velocities, \vc, were initially estimated from the orbital period, $P$\ \citep{vil21} and the stellar radius, $R$, assuming that the orbital was circular; we return to this assumption below. Then

\begin{equation} \label{eq_vs}
  \varv_{\rm{c}}= \frac{2\pi R}{P}
\end{equation}

In turn the stellar radius was deduced from estimates of the stellar luminosity, $L$ and the effective temperature, \teff \cite[see, for example,][]{gra05}

\begin{equation} \label{eq_r}
  \frac{R}{R_{\odot}}=\sqrt{\frac{L}{L_{\odot}}\left (\frac{T_{\odot}}{T_{\mathrm{eff}}}\right )^4}
\end{equation} 

\noindent where \Lsun, \Rsun\ and \Tsun\ are the solar luminosity, radius and effective temperature respectively.

Effective temperatures were already available for 36 SB1/\sbs\  primaries \citep{mce15, gar17}. These have been derived from the silicon ionisation equilibrium and/or the \ion{He}{ii} spectrum and utilised  {\sc tlusty} non-LTE model atmosphere grids \citep{hub88, hub95, lan07, rya03, duf05}. The spectroscopy of the remaining 33 systems was inspected to identify possible effective temperature estimators. The \ion{He}{ii} spectrum was identified in 10 targets and the \ion{Si}{iii} and \ion{Si}{iv} lines in one target (\#821).  The former were used to estimate effective temperatures, with the latter yielding an estimate approximately 800\,K lower than that from the \ion{He}{ii} spectrum. The methodology adopted was as discussed by \citet{mce15}, \citet{gar17} and \citet{duf19} where more details can be found. For the remaining 25 targets, effective temperatures for the primary were estimated using the \teff--spectral-type calibration of \citet{tru07} for the LMC. All these estimates are summarised in Table \ref{t_SB1_full}.

\citet{mce15} and \citet{gar17} estimated stochastic errors in their effective temperature estimates of $\pm$1000\,K and this is consistent with the quality of the \ion{He}{ii} fits for the 10 targets analysed here. \citet{duf19} discussed the stochastic errors associated with using the spectral type calibration of \citet{tru07} and estimated a typical uncertainty of $\pm$1500\,K. Comparison of the effective temperature estimates from the spectral type compared with those from the silicon ionisation equilibrium or \ion{He}{ii} lines yielded a mean percentage difference of 800$\pm$1100\,K or 3.3$\pm$4.9\%\footnote{For some targets, \citet{gar17} could only deduce a range of possible effective temperature estimates and these have been excluded from the comparison.} Here we adopt an error estimate of 10\%  for all the \teff-estimates.

Luminosity estimates for our targets were obtained from the estimates of their apparent magnitudes, extinction and bolometric correction adopting a distance modulus of 18.49$\pm$0.09 \citep{deG14}. Intrinsic (B-V) colours were estimated using the (B-V)--\teff\ calibration of \citet{cas03} for a metallicity 0.5\,dex lower than the sun and combined with the photometry discussed in \citet{eva11} to estimate reddening, E(B-V). \citet{mai14} discussed the extinction of some of the VFTS targets and obtained individual extinction curves but here we follow the approach of \cite{dor13} and \citet{mce15} and adopt a single values for the ratio of the V-band extinction to E(B-V) of 3.5$\pm$1.0. As discussed by these authors, this should be adequate for our targets that exhibit low to medium amounts of interstellar extinction. Bolometric corrections were then taken from \citet{ped20} and combined with the apparent V magnitude, extinction and distance modulus to provide estimates of the absolute bolometric magnitudes and luminosities (in terms of the solar luminosity). The latter are summarised in Table \ref{t_SB1_full}, together with the adopted photometry and estimated reddening.

Stochastic uncertainties in the bolometric magnitudes will arise from several sources, viz.:
\begin{enumerate}
    \item the distance modulus, where we adopt the error estimate of \citet{deG14}.
    \item the intrinsic colours, where an error of 10\% in the effective temperature estimate translates in to a typical error of $\pm$0.02 in E(B-V) and  $\pm$0.07 in A$_{\rm{v}}$.
    \item the ratio of the extinction to E(B-V). The median E(B-V) for the SB1/\sbs sample (see Table \ref{t_SB1_full}) is 0.32, which would translate into a comparable error in the extinction.
    \item the bolometric correction, where an error of 10\% in the effective temperature estimate lead to a typical error of 0.25 in the bolometric magnitude.
\end{enumerate}
There will be additional uncertainties in the observed photometry but as discussed by \citet{eva11}, these will be small compared with those listed above. Combining the above errors in quadrature leads to a typical uncertainty of 0.42 in the bolometric magnitude or 0.17 dex in \logl. 

The uncertainties in the effective temperature and luminosity will in turn determine the reliability of the estimates of the stellar radius, $R$. These are related as shown in Equation \ref{eq_r}, which implies typical uncertainties in the stellar radius of 20\% from the \teff-uncertainty and 22\% from the \logl-uncertainty. Again combining these in quadrature leads to a total uncertainty in the $R$-estimates of approximately 30\%.

Synchronous velocities, \vc, have been estimated from Equation \ref{eq_vs} and are also listed in Table \ref{t_SB1_full}. The periods deduced by \citet{vil21} normally have an accuracy of better than 1\% with the largest uncertainty being 2.3\%. Hence the uncertainty in \vc-estimate will be dominated by that in the radius leading us to adopt a typical uncertainty of 30\%. We have calculated the ratios, \fci, of the vsini-estimates to the \vc-estimates and list these in Table \ref{t_SB1_full}. As they cover a wide range, we will base the discussion on their logarithm, \logfc, with synchronous rotation only being possible if \logfc$\la$0. The uncertainty in the \vc-estimates translate into an uncertainty of 0.13\,dex in \logfc. Combining this with the uncertainties in the \vsini-estimates discussed above lead to uncertainties in the range 0.13--0.16\,dex and here we adopt a typical {\em stochastic} uncertainty of $\pm$0.15\,dex in \logfc.

The above approach has not included any contribution of the secondary to the estimated luminosity. As such the primary luminosity may be systematically overestimated and in turn this may lead to an overestimation in the \vc-values and an underestimation of the \fci-values. As examples, a 20\% and 40\% secondary contribution to the observed flux would lead to an increase in the \fci-value of approximately 11 and 23\% respectively, translating to an increases of 0.05 and 0.11\,dex in \logfc. These will be discussed further in Sect.\ \ref{s_disc}.

The initial \fci-estimates have assumed that the eccentricity is zero and as such the synchronous velocity does not vary with position in the orbit. Following the methodology of \citet{hut81}, we have also estimated the instantaneous synchronous velocity at periastron, \vp , and the corresponding $f$-ratios, which we designate as \fpe\ in order to emphasise that it relates to a stellar component when at periastron. These ratios are also summarised in Table \ref{t_SB1_full}. As discussed by \citet{hut81}, it is expected that the stellar components will firstly become pseudo-synchronised at periastron. Also these ratios provide the strongest constraint on a system being asynchronous and hence they should allow an upper limit to be found for the frequency of synchronous systems. The effect of a significant light contribution from the unseen companion and the stochastic uncertainties in the \logfp-values will be similar to those for the \logfc-values, as discussed above.

\section{SB2 sample}
\subsection{Projected rotational velocities, \vsini} \label{s_vsini_sb2}
Projected rotational velocities for some of the primaries in the SB2 systems have been previously estimated by \citet{gar17}, who used a subset of the VFTS exposures that had been obtained consecutively at effectively the same epoch. We have used the more extensive BBC spectroscopy to attempt to estimate the rotational broadening for both binary components. Our procedure was to identify exposures where the two components had significantly different radial velocities (which were normally the exposures that had been used by \cite{vil21} to constrain the orbital solutions) and then to shift these to the rest-frame of the primary. The resulting combined spectrum consisted of the primary contaminated by the secondary spectrum that had been 'smeared out' due to the varying radial velocity separation of the two components. A similar procedure was undertaken for the secondary.

For all 14 systems, \vsini-estimates could be obtained for the primary. Generally these were obtained using the same Fourier Transform (FT) methodology as for the SB1 system discussed in Sect.\ \ref{s_vsini_sb1}. For two targets, (\#112 and \#520), the \ion{O}{ii} multiplet near 4070\AA\ was also used and due to line-blending, profile fitting (PF) results were adopted. A comparison of the primary \vsini-estimates using the different methodologies yielded a mean difference of -0.4$\pm$10.2\,\kms\ (PF estimate - FT estimate), implying that the use of different methodologies should not introduce significant additional uncertainties. Reliable \vsini-estimates of the secondary could only be obtained for 5 systems -- see Appendix \ref{a_sb2} for details and additional marginal measurements. This reflected the greater spectral contamination due to the brighter primary. In Table \ref{t_vsini_sb2}, the mean values of the \vsini-estimates are summarised together with their sample standard deviations and the number of estimates.

The combined spectra were also used to re-assess the spectral types. For the primaries, these were found to be in good agreement with those of \citet{eva15} with differences being at most one half of a subclass. Spectral classification was only possible for 6 secondaries and should be treated with caution especially in regard to the luminosity class. However they were consistent with the secondary being the fainter member of the binary and with the estimated flux ratios, $F$, discussed in Sect.\ \ref{s_f_sb2}. The new spectral types and those of \citet{eva15} are also summarised in Table \ref{t_vsini_sb2}.

\begin{table*}
\caption{\vsini-estimates for primaries and secondaries in the SB2 sample denoted by subscripts 1 and 2, together with $\sigma$-estimates and the number of lines ($n$) used for these determinations. Spectral types for primaries from \cite{eva15} are listed, as are those from the current work for both components where available.}\label{t_vsini_sb2}
\begin{center}
\begin{tabular}{llllrrrrrrrrrrr} 
\hline\hline
VFTS &   \multicolumn{3}{c}{Spectral Types}   &   ~~~~~~\vsini$_1$ &   $\sigma_1$    &   $n_1$ 
&   ~~~~~~\vsini$_2$ &   $\sigma_2$    &   $n_2$ \\ 
    & Evans         &   Primary   &  Secondary& \kms   &       &       & \kms   &       &       \\ \hline
112 & Early-B       &   B0.5 V    &   B1: V:  &   141 &   17  &   6   &   109 &   40  &   4   \\
199 & Early-B       &   B2 V      &   B2: V:  &   98  &   14  &   4   &   81  &   7   &   4   \\
206 & B3 III:       &   B3 III-V  &   --      &   56  &   7   &   4   &   --  &   --  &   --  \\
240 & B1-2 V        &   B1: V     &   --      &   141 &   16  &   2   &   --  &   --  &   --  \\
248 & B2: V         &   B1.5 V    &   --      &   96  &   27  &   6   &   --  &   --  &   --  \\
255 & B2: V         &   B2 V      &   B2: V:  &   67  &   16  &   5   &   59  &   14  &   3   \\
364 & B2.5: V       &   B2.5 V    &   --      &   96  &   8   &   5   &   --  &   --  &   --  \\
383 & B0.5: V       &   B0.5 V    &   --      &   61  &   8   &   6   &   --  &   --  &   --  \\
520 & B1: V         &   B1 V      &   B1.5: V:&   81  &   8   &   7   &   --  &   --  &   --  \\
589 & B0.5 V        &   B0.5 V    &   B1: V:  &   <40 &   --  &   --  &   <40 &   --  &       \\
637 & B1-2 V+EarlyB &   B1.5 V    &   --      &   138 &   9   &   4   &   --  &   --  &   --  \\
686 & B0.7 III      &   B0.7 III  &   --      &   <40 &   --  &       &   --  &   --  &   --  \\
752 & B2 V          &   B1.5 V    &   B1.5: V:&   70  &   7   &   6   &   67  &   3   &   3   \\
883 & B0.5 V        &   B0.5 V   &   --       &   96  &   16  &   3   &   --  &   --  &   --  \\
\hline
\end{tabular}
\end{center}
\end{table*}

\subsection{Synchronous velocities}\label{s_f_sb2}

The determination of the synchronous velocities followed the procedures used for the SB1 targets discussed in Sect. \ref{s_f_sb1} but the SB2 classification led to additional complications. Firstly when estimating the bolometric correction, the effective temperature estimate for the primary (using the same spectral type calibration as for the SB1 sample) was adopted. When the secondary makes a significant contribution to the observed visible flux, it would be expected to have a similar (but normally smaller) effective temperature. Hence we do not expect that this will lead to significant additional uncertainties and adopt the typical stochastic uncertainty found for the SB1 targets, of $\pm$0.17 dex for the logarithmic luminosity of the binary. Table \ref{t_f_sb2} summarises the photometry, effective temperatures, bolometric corrections and total luminosities estimated for our SB2 sample.

To estimate the ratio ($F$) of the luminosity of the two components, we have used the ratio of the stellar masses found from the orbital solutions and then assumed that the stellar luminosity is proportional to the mass raised to the power $\alpha$, viz.

\begin{equation}
F=\left(\frac{M_p}{M_s}\right)^{\alpha}= \left(\frac{K_s}{K_p}\right)^{\alpha}
\end{equation}
\noindent where $K_p$\ and $K_s$\ are the radial velocity semi-amplitudes of the primary (defined as the component with the smaller value of $K_p$) and secondary respectively and have been taken from \citet{vil21}.

There have been numerous studies of the mass-luminosity relationship \citep[see, for example,][]{wan18, eke15, dem91, gri88} that have found both different $\alpha$-values and that $\alpha$ may vary with stellar mass. For our B-type targets that have masses in the range 7--15\Msun\ \citep{sch18}, values in the range $\sim$3--4 have been found and here we adopt $\alpha$=3.5$\pm$0.5. The corresponding $F$-values are summarized in Table \ref{t_f_sb2}\footnote{Two system, \#206 and \#686, have a giant luminosity classification for the primary and hence this procedure may be unreliable.}.

Using a similar procedure as for the SB1 sample, it was then possible to estimate the stellar radii, synchronous velocities and $f$-values\footnote{when either `$f$-value' (or its logarithm) is used, the discussion applies to both \fci-values and \fpe-values} for all the primaries and the 5 secondaries that had \vsini-estimates; these estimates are summarised in Table \ref{t_f_sb2}. Effective temperature estimates appropriate to their spectral type were adopted and those for the primaries are listed in Table \ref{t_f_sb2}. No estimates are listed for the secondaries but these can be inferred from estimates adopted for primaries with the same spectral type. 

The error budget for the synchronous velocities will include those uncertainties discussed in Sect.\ \ref{s_f_sb1} but there may be additional sources -- for example, the division of the total luminosity between the two components.  A plausible range for the flux contribution of the primary is 50\% to 84\% -- the former being constrained by its primary designation and the latter from requiring a flux ratio, $F\leq 5$ to allow the identification of both components. This is consistent with the estimated values for our sample that range from 53\% to 83\% of the total flux. For a median primary flux contribution off 63\% this would then translate into maximum uncertainties of approximately 15\% in the synchronous velocities. Combining this estimated uncertainty in quadrature with those discussed in Sect.\ \ref{s_f_sb1} would then lead to a total uncertainty of approximately 33\% in these estimates.  Including the uncertainty in the  \vsini-estimates lead to a maximum error budget for the \logf-values of approximately $\pm$0.16\,dex.

The uncertainties for the $f$-values of the secondaries may be larger due to the uncertainty in the spectral types (and hence effective temperature) and the greater sensitivity of the synchronous velocities to uncertainties in the flux distribution between the components. The secondaries have estimated spectral types of B1 V to B2 V. For a spectral type of B1.5 V, our adopted uncertainty in the \teff-estimate of $\pm$10\% would correspond to a spectral type range of approximately B0.5 to B2.5 V. Hence despite the additional uncertainty in the secondary's spectral type, we have retained this error estimate. The flux contributions of the secondaries range from 17\% to 47\% and following the same methodology as for the primaries leads to a maximum uncertainty in the $R$-estimates of approximately 45\%. Including uncertainties in the \vsini-estimates then leads a maximum error budget for the \logf-values of approximately $\pm$0.18\,dex. Here we adopt a conservative error estimate of $\pm$0.20\,dex for all the \logf-estimates for the SB2 sample.

\begin{table*}
\caption{Observed and derived parameters for the primaries in the SB2 sample, where \vc\ is the synchronous velocity as defined by equation\,1, with \logfc and \logfp being the inferred ratios of \vsini\ to \vc for the cases of circular and eccentric orbits respectively. Columns headed `1' and `2' refer to primary and secondary components. $F$ is the flux ratio from equation 3, and $R$ denotes the stellar radii.}\label{t_f_sb2}
{\tiny
\begin{center}
\begin{tabular}{lrrrrrrrrrrrrrrrr}
\hline\hline
VFTS & V     &  (B-V) & \teff &  E(B-V)& BC    & log\,$L/L_{\odot}$  & $F$  & $P$   &\multicolumn{2}{c}{~~~~~$R$(\Rsun)}  & \multicolumn{2}{c}{\vc(\kms)} & \multicolumn{2}{c}{~~~~~\logfc} & \multicolumn{2}{c}{~~~~~\logfp} \\
&&&1000\,K&&&&&d&1&2&1&2&1&2&1&2\\ \hline
112 & 16.19 &  0.08 & 29.1 & 0.35 & -2.85 & 4.45 & 1.89 &  1.67 &  5.4 & 4.6 & 162 & 139 & -0.06       & -0.11 & -0.06 & -0.11 \\
199 & 16.91 & -0.01 & 24.5 & 0.23 & -2.43 & 3.83 & 1.18 &  1.67 &  3.3 & 3.1 & 102 &  94 & -0.02       & -0.06 & -0.02 & -0.06 \\
206 & 14.99 &  0.15 & 19.2 & 0.33 & -1.84 & 4.51 & 1.48 &  7.56 & 12.5 &     &  84 &     & -0.18       &       & -0.20          \\
240 & 15.85 &  0.01 & 26.8 & 0.26 & -2.65 & 4.39 & 1.12 &  1.38 &  5.3 &     & 195 &     & -0.13       &       & -0.22          \\
248 & 16.49 &  0.00 & 25.7 & 0.25 & -2.55 & 4.07 & 2.59 &  2.49 &  4.6 &     &  94 &     &  0.01       &       & -0.14          \\
255 & 16.68 &  0.02 & 24.5 & 0.26 & -2.43 & 3.96 & 1.12 &  2.84 &  3.9 & 3.7 &  69 & 65  & -0.01       & -0.04 & -0.03 & -0.06 \\
364 & 16.82 &  0.03 & 23.4 & 0.26 & -2.32 & 3.86 & 1.11 &  4.40 &  3.8 &     &  43 &     &  0.35       &       &  0.20          \\
383 & 16.10 &  0.20 & 29.1 & 0.47 & -2.85 & 4.66 & 3.48 &  2.60 &  7.4 &     & 144 &     & -0.37       &        & -0.43         \\
520 & 16.69 &  0.08 & 26.8 & 0.33 & -2.28 & 3.96 & 4.56 &  2.95 &  5.0 & 2.7 &  85 & 46  & -0.02       &        & -0.15          \\
589 & 15.83 &  0.15 & 29.1 & 0.82 & -2.85 & 4.69 & 2.19 &  7.63 &  7.3 & 5.8 &  48 & 38  & $\leq$-0.08 & $\leq$0.02 
& $\leq$-0.34 & $\leq$-0.24 \\
637 & 16.61 &  0.04 & 25.7 & 0.29 & -2.55 & 4.15 & 1.53 &  1.63 &  4.3 &     & 133 &     &  0.01       &        & -0.01          \\
686 & 14.99 &  0.17 & 25.4 & 0.41 & -2.52 & 4.89 & 3.45 & 16.87 & 12.8 &     &  38 &     & $\leq$0.02  &        & $\leq$-0.28    \\
752 & 16.48 & -0.12 & 25.7 & 0.13 & -2.55 & 3.90 & 1.40 &  1.41 &  3.5 & 2.9 & 124 & 105 & -0.25       & -0.20  & -0.25 & -0.20 \\
883 & 16.49 &  0.13 & 29.1 & 0.40 & -2.85 & 4.40 & 2.70 &  3.50 &  5.4 &     &  78 &     &  0.09       &        & -0.26          \\ \hline
\end{tabular}
\end{center}
}
\end{table*}

\section{Discussion}\label{s_disc}

\subsection{Supergiant primaries}
Our sample consists of mainly luminosity class III to V primaries with only 7 SB1 supergiant primaries; these are briefly discussed first. Their orbital periods range from approximately 9 days to more than 200 days. The $f$-values for the two Ia supergiants (\#430 and 576) are consistent with synchronous rotation, whilst those for the other Ib/II supergiants are not, although including the error estimates discussed above would lead to two additional objects (\#687 and \#827) fulfilling the criterion. However given the very small sample, the only firm conclusion is that the remaining three targets (\#291, \#779, \#829) that are all luminosity class II or II-Ib cannot be synchronised, having \logfc values significantly greater than zero (0.33, 0.59 and 1.40 respectively).

\subsection{Orbital period constraints} \label{d_per}
\begin{figure}
	\includegraphics[width=8.5cm, angle=0]{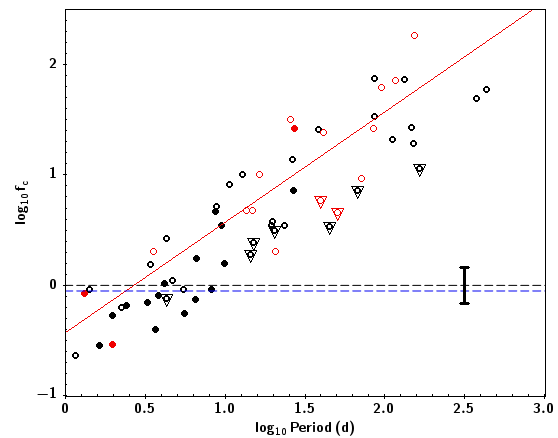} \\
	\includegraphics[width=8.5cm, angle=0]{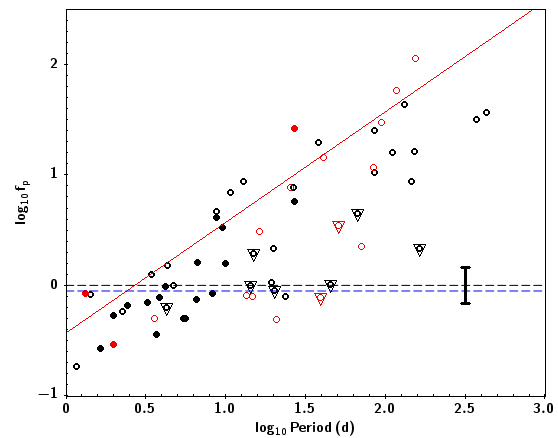}
	
	\caption{Plots of \logfc- and \logfp-values for the luminosity class III--V primaries in the SB1 (black symbols) and \sbs\ (red symbols) systems. Upper limits are indicated as downward triangles. Filled symbols denote systems that have eccentricities within 2-$\sigma$ of zero. Also shown are the upper limits for synchronous rotation assuming that the secondaries contribute negligible flux (dashed black line) and 20\% (dashed blue line) of the total flux. The red dashed line in the upper panel shows the position at period P that a `typical' B-type star with radius 8\,\Rsun and \vsini=150\,\kms would have for synchronisation. Also shown is the typical error estimate.}
	\label{f_SB1}
\end{figure}

\begin{figure}
	\hspace{-10pt}
	\includegraphics[width=8.5cm, angle=0]{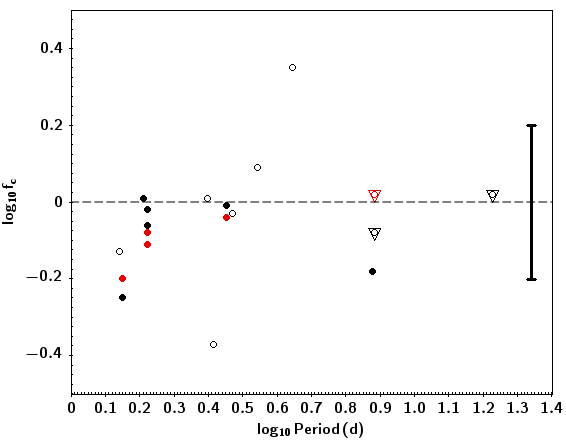} \\
	\includegraphics[width=8.5cm, angle=0]{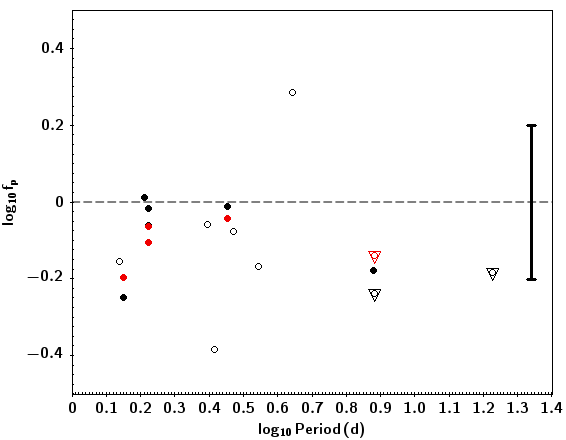}
	\caption{Plots of \logfc- and \logfp-values for the primaries (black symbols) and secondaries (red symbols) in the SB2 systems. Upper limits are indicated as downward triangles. Filled symbols denotes systems that have eccentricities within 2-$\sigma$ of zero. Also shown are a typical error estimate and the upper limit for synchronous rotation (dashed black line). Estimates for \#199 have been slightly shifted to aid legibility.}
	\label{f_SB2}
\end{figure}

Estimates of the \logf-values for both our SB1\footnote{Henceforth discussion of the SB1 sample will include both the SB1 and \sbs\ targets unless otherwise specified} and SB2 samples with luminosity class III to V primaries are plotted as a function of orbital period in Fig. \ref{f_SB1} and \ref{f_SB2} respectively. The upper limit for synchronisation is shown, together with that assuming that the SB1 secondaries contribute 20\% of the total flux. Fig.\ \ref{f_SB1} implies that a significant fraction of the SB1 systems with $\log P\geq0.5$\ cannot be synchronised. Indeed excluding upper limits for the $f$-values, effectively all the systems with $\log P\geq1.0$\ must be un-synchronised. These conclusions remain robust after allowing for a flux contributions from the secondary and the estimated stochastic errors; it is also unaffected by which set of $f$-values is adopted. By contrast all systems with $\log P < 0.5$\ could be synchronised. Similar results are found for the SB2 systems (see Fig.\ \ref{f_SB2}), although now only three actual $f$-estimates (rather than upper limits) are available for $\log P\geq0.5$.

Recalling that sin$i$ is in principle unknown, a short period system may not actually be synchronised despite having an $f$-value less than unity.
For a given stellar radius the synchronisation velocity will be inversely proportional to the period -- see Fig.\ 3 of \citet{lev76}. As such the criterion for synchronisation becomes less demanding as the period decreases. For example, a star with a typical radius of 7\Rsun\ in a system with an orbital period, $P$=3d, only requires \vsini$\la$120\kms\ to fulfil the synchronisation criterion. 

\begin{figure}
	\hspace{-10pt}
	\includegraphics[width=8.5cm, angle=0]{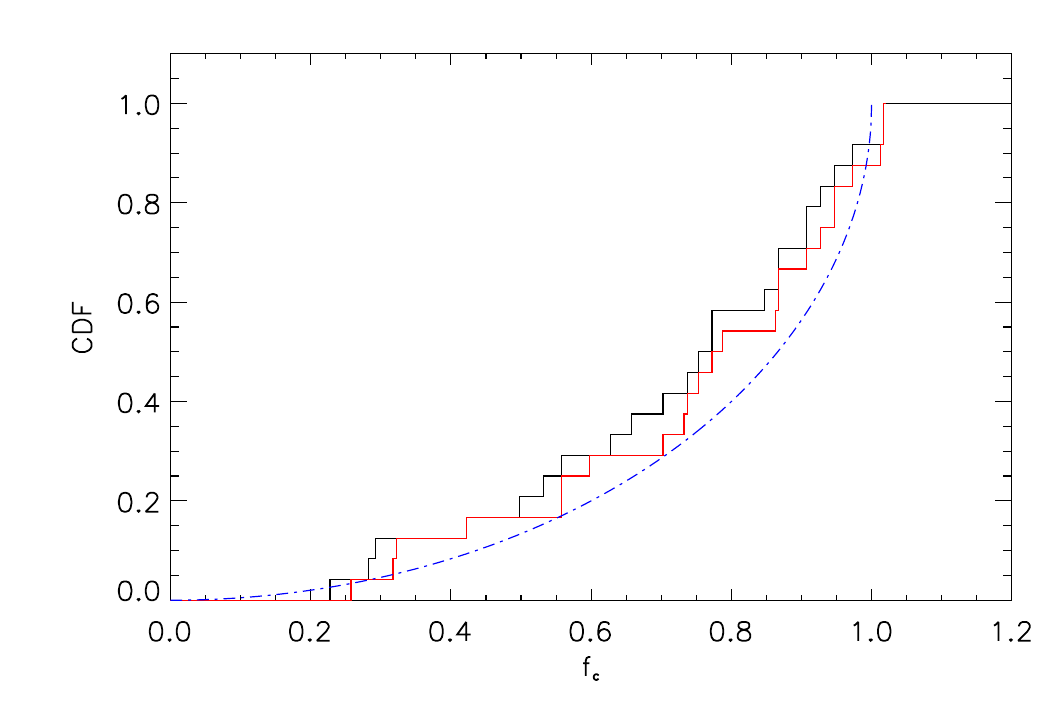} \\
	\includegraphics[width=8.5cm, angle=0]{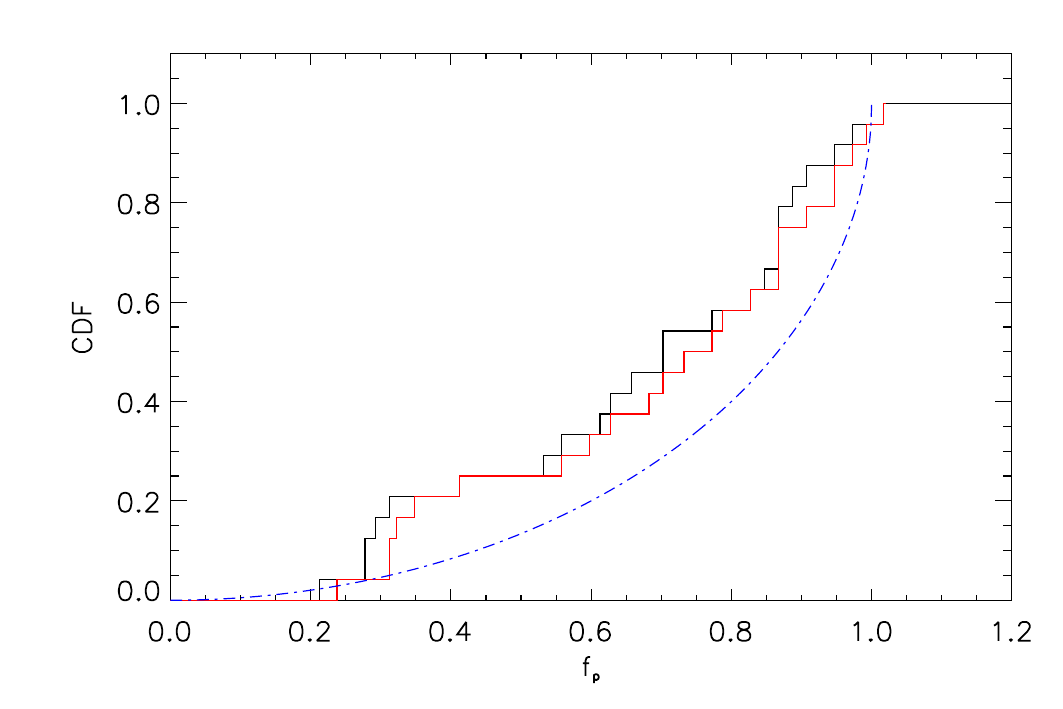} 
	\caption{Cumulative distribution functions for the \logfc- and \logfp-values for systems with $\log P<3$\,days. The black curves assume that the SB1 secondaries contribute negligible flux, whilst the red curves assume that they contribute 20\% of the total flux. The dotted blue curves are the expected behaviour for synchronous rotation, alignment of the rotational and orbital axes and a random distribution of orbital axes.}
	\label{f_cdf}
\end{figure}

In Fig.\ \ref{f_cdf}, the cumulative distribution function (CDF) has been plotted for our two sets of $f$-values for all the SB1 and SB2 systems with $\log P < 0.5$. Two distributions are shown corresponding to the SB1 secondaries contributing either negligible flux or 20\% of the total flux. If the systems were synchronised and the orbital and rotational axes were aligned, the $f$-values would become the sine of the rotational angle ($i$) of inclination. Therefore we also plot the CDF for $\sin i$ assuming that the inclination angles are randomly distributed \citep[see, for example,][]{gra05}. The agreement between the observed and theoretical CDFs is reasonable for the \fci-estimates but poorer for the \fpe-estimates.

\begin{table}
\caption{Results of Kolmogorov-Smirnov (KS) and Kuiper statistical tests on the different $f$-estimates. The final column lists the probability that their CDFs follow that predicted for a random distribution of rotational axes.}\label{t_KS}
\begin{center}
\begin{tabular}{lcrrr} 
\hline\hline
Sample  &  Secondary & \multicolumn{2}{c}{Probability} \\ 
& contribution & KS & Kuiper\\ \hline
\fci-estimates  & 0\%  & 19.9\% & 19.0\%\\
\fci-estimates  & 20\% & 50.1\% & 22.6\%\\
\fpe-estimates  & 0\%  & 2.3\%  &  5.2\%\\
\fpe-estimates  & 20\% & 8.3\%  & 20.3\%\\
\hline
\end{tabular}
\end{center}
\end{table}

Kolmogorov-Smirnov  \citep{fas87} and Kuiper \citep{mod11} statistical tests were performed on the different samples and are summarised in Table \ref{t_KS}. In general they support the above comments. For the \fci-estimates, relative high probabilities are found for them being drawn from a population with a random distribution of rotational axes. By contrast, a range of probabilities are found for the \fpe-estimates from less than 5\% to 20\%.

The above analysis assumes that all system with periods, $\log P < 0.5$, are synchronised. Additionally the inclination angles of the orbital (and hence rotational) motion may not be random. Systems with larger orbital inclination angles would have larger variations in their observed radial velocities and might have been preferentially identified. In fact the reverse is seen with the CDFs, particularly for the \fpe-estimates, implying that systems with small inclination angles may be more numerous than would be expected from a random distribution of angles of inclination.

In summary, most systems with periods, $\log P < 0.5$, may be synchronised, whilst effectively all system with periods, $\log P > 1.0$, cannot be synchronised. For the shorter period systems, their $f$-estimates are consistent with those expected from a random distribution of orbital inclination axes, with the estimates favouring synchronisation at the mean orbital velocity rather than that at periastron.

\subsection{Separation constraints} \label{d_sep}
The variation of the degree of synchronisation with the separation of the binary components (excluding the supergiant primaries) has also been investigated. We will consider the fractional stellar radius ($r$), which we define as the stellar radius divided by the sum of the semi-major axes of the binary orbit, $a$, \citep[see, for example][]{max20}. For the SB2 systems the semi major axes can be inferred from the orbital parameters, provided the angle of orbital inclination is known. We will assume that all the SB2 systems with $f$-values less than one are synchronised and then $\sin i=f$\footnote{Three systems with 0.0$<$\logfc$<$0.1 have been included with $\sin i$\ set to unity}. The $r$-estimates are not significantly affected by our choice of $f$-values and here we have chosen to use the \fci-estimates and summarise the results in Table \ref{t_a_SB2}. We also tabulate stellar masses derived from the $M\sin^3 i$-estimates of \citet{vil21}.

The stellar mass estimates range from approximately 3 to 20\Msun\ and have been compared with those of \citet{pec13} for the luminosity V classifications. For eight systems (\#112, 248, 255, 383, 589, 637, 752, 883) the agreement is reasonable with the mass estimates disagreeing by less than 30\%. For three systems (\#199, 240, 520) agreement is poorer with differences of up to a factor of three. However our mass estimates are very sensitive to the errors in \logfc\ due to the $\sin^3 i$\ dependence of the results taken from \citet{vil21}. In turn given the adopted error in \logfc\ of $\pm$0.2\,dex, such differences are not surprising.

Estimates of $r$ vary from 0.14 to 0.50 with a median value of 0.26\footnote{Use of the \fpe-estimates would have lead to a range in $r$ of 0.11 to 0.47 with a median of 0.25}. For the SB2 systems, we can also estimate the effective radii of the Roche lobes (\rl)  using the methodology of \cite{egg83}. These are also summarised in Table \ref{t_a_SB2}, together with the ratios (\fl) of the estimated stellar radius to that of the Roche Lobe:

\begin{equation}
    f_L=\frac{R}{R_{\rm L}}
\end{equation}

\noindent The \fl-estimates are relatively large and imply that most of the SB2 stars occupy a significant amount of their Roche Lobe. VFTS\,240 has an \fl-estimate greater than unity but as discussed below for the SB1 systems, the uncertainties are consistent with \fl$\leq 1$. We discuss the $r$ and \fl-estimates further below after consideration of the SB1 systems.

\begin{table*}
\caption{The estimated semi-major axes ($a$) and fractional stellar radii ($r$) for the primaries (columns headed `1') and secondaries (`2') in the SB2 sample, that appear to be synchronised. Also listed are estimated Roche radii (\rl) and the ratios to the stellar radii to their Roche lobs radii (\fl). Columns 'Mass' are stellar masses (\Rsun) derived here from the orbital solution \citep{vil21} and sin\,$i$, while 'ST Mass' is the corresponding estimates inferred from a mass -- spectral type calibration \citep{pec13}. }\label{t_a_SB2}
\begin{center}
\begin{tabular}{lrrrrcrrrrrrr}
\hline\hline
VFTS & \multicolumn{2}{c}{~~~~~Mass} & \multicolumn{2}{c}{~~~ST Mass} & $a$(\Rsun)  &\multicolumn{2}{c}{~~~~~~$r$}  & \multicolumn{2}{c}{~~~~~\rl(\Rsun)} & \multicolumn{2}{c}{~~~~~\fl}\\
&1&2&1&2&&1&2&1&2&1&2\\ \hline
112 & 17 & 14 & 15 & 12 & 18.6 &  0.29  &  0.25 &   7.36  &  6.77  &  0.73  &  0.68 \\
199 & 4  & 4  &  7 &  7 & 11.7 &  0.29  &  0.26 &   4.47  &  4.37  &  0.75  &  0.71 \\
206 & 12 & 11 &    &    & 46.1 &  0.27  &       &  17.40  &        &  0.72  &       \\
240 &  4 &  4 & 12 &    & 10.6 &  0.50  &       &   3.66  &        &  1.45  &       \\
248 & 10 &  8 & 10 &    & 20.1 &  0.23  &       &   6.74  &        &  0.69  &       \\
255 & 10 & 10 &  7 &  7 & 22.7 &  0.17  &  0.16 &   8.46  &  8.34  &  0.46  &  0.44 \\
383 & 17 & 12 & 15 &    & 24.2 &  0.31  &       &   9.32  &        &  0.79  &       \\
520 &  5 &  3 & 12 & 10 & 17.9 &  0.28  &       &   6.40  &        &  0.78  &       \\
589 &  7 &  4 & 15 & 12 & 38.3 &  0.19  &  0.15 &   11.19 &  10.10 &  0.65  &  0.57 \\
637 &  8 &  7 & 10 &    & 14.7 &  0.29  &       &   5.57  &        &  0.77  &       \\
686 &  5 &  3 &    &    & 55.0 &  0.23  &       &  15.53  &        &  0.82  &       \\
752 &  8 &  8 & 10 & 19 & 13.3 &  0.26  &  0.22 &   5.14  &  4.91  &  0.67  &  0.59 \\
883 & 11 &  8 & 15 &  4 & 25.6 &  0.21  &       &   6.64  &        &  0.81  &       \\
\hline
\end{tabular}
\end{center}
\end{table*}

We can also estimate semi-major axes for our SB1 systems although now the process is more complicated. We again assume that all the SB1 system with $f$-values less than unity are synchronised. Using the implied inclination angle, the mass of the secondary can be inferred from the mass function provided that of the primary is known. Then the mass ratio can be used to estimate the semi-major axis for the secondary and hence the semi major axis for the system.

Mass estimates for the primaries have been estimated using {\sc bonnsai}\footnote{The {\sc bonnsai} web-service is available at: \newline \url{www.astro.uni-bonn.de/stars/bonnsai}.}, which utilises a Bayesian methodology and the grids of models from \citet{bro11a} to constrain the evolutionary status of a given star \citep[see][for details]{sch14}. As independent prior functions, we adopted the LMC metallicity grid of models, a \citet{sal55} initial mass function, the initial rotational velocity distribution estimated by \citet{duf12}, a random orientation of spin axes, and a uniform age distribution. The estimates of effective temperature, luminosity and projected rotational velocity (taken from Table~\ref{t_SB1_full}) were then used to constrain current primary masses ($M_1$). 

Using the errors estimates from  Sect.\ \ref{s_vsini_sb1} for the effective temperature, luminosity and projected rotational velocities, {\sc bonnsai} returns 1$\sigma$-uncertainties for all the quantities that it estimates. In the case of the stellar masses, these were generally 11-13\% and never greater than 14\%. An additional systematic uncertainty may arise from our implicit assumption that the secondary makes no contribution to the optical flux. Additional simulations showed that a secondary contribution of 20\% of the total flux would decrease the mass estimates by 4-6\%.

\begin{table}
\caption{The estimated primary ($M_1$) and secondary masses ($M_2$),  semi-major axes ($a$) (units are \Msun\ and \Rsun), and fractional stellar radii ($r$) for the primaries in the SB1 sample that appear to be synchronised. Also listed are their estimated primary Roche lobe radii (\rl\ in \Rsun) and the ratios to the stellar radii to their Roche lobe radii (\fl).}\label{t_a_SB1}
\begin{center}
\begin{tabular}{llrrrrrrr}
\hline\hline
VFTS & $P$(d) & $M_1$ & $M_2$ & $a$  &$r$  &\rl & \fl\\ \hline
009 & 4.71 & 12.0 & 2.5 & 28.8 & 0.26 & 14.3 & 0.52 \\
027 & 6.58 & 12.2 & 6.7 & 39.3 & 0.41 & 17.0 & 0.95 \\
033 & 3.86 & 10.4 & 1.8 & 23.8 & 0.30 & 12.4 & 0.58 \\
179 & 1.16 & 10.6 & 1.7 & 10.7 & 0.47 &  5.1 & 0.97 \\
189 & 1.43 & 13.0 & 3.4 & 13.6 & 0.49 &  6.5 & 1.02 \\
225 & 8.24 & 12.8 & 1.9 & 42.1 & 0.25 & 22.3 & 0.47 \\
246 & 2.44 & 12.2 & 2.8 & 18.8 & 0.53 &  9.6 & 1.04 \\
305 & 4.18 &  8.8 & 2.1 & 24.2 & 0.19 & 12.0 & 0.38 \\
324 & 1.64 & 13.0 & 3.7 & 15.0 & 0.43 &  7.1 & 0.91 \\
342 & 4.28 &  7.8 & 1.4 & 23.2 & 0.19 & 11.2 & 0.39 \\
434 & 5.61 & 10.6 & 2.2 & 31.0 & 0.28 & 15.2 & 0.58 \\
534 & 3.69 & 17.0 & 6.3 & 28.3 & 0.37 & 12.7 & 0.82 \\
662 & 1.99 &  7.2 & 0.6 & 13.2 & 0.68 &  7.8 & 1.15 \\
705 & 2.26 & 11.0 & 5.2 & 18.3 & 0.34 &  7.8 & 0.80 \\
730 & 1.33 & 12.0 & 0.3 & 11.7 & 0.65 &  8.0 & 0.96 \\
788 & 3.26 & 10.6 & 3.3 & 22.2 & 0.34 & 10.7 & 0.71 \\
888 & 1.97 & 11.0 & 3.8 & 16.2 & 0.34 &  7.7 & 0.71 \\
891 & 5.48 &  9.2 & 0.6 & 28.0 & 0.23 & 12.5 & 0.52 \\\hline
\end{tabular}
\end{center}
\end{table}

Table \ref{t_a_SB1} summarises the estimates for the semi-major axes, fractional radii and effective Roche Lobe radii for the SB1 systems that have $f$-values less than unity and may be synchronised. We have adopted the \fci-estimates, which leads to 18 systems, including two (\#342, \#799) that have upper limits to their \vsini-estimates; for the latter, we adopted the angle of inclination implied by these limits. We have also included 2 targets (\#009, \#305) that have \fci-estimates within $1\sigma$\ of being synchronised, that is \logfc$\leq$0.15.

The adoption of the \fpe-estimates would have led to six additional systems (\#213, 218, 388, 665, 784, 799) fulfilling the criterion for synchronisation. These have large eccentricities, $e>0.5$, and additionally five systems are classified as \sbs. Given the uncertainty about their designation as single line spectroscopic binaries implied by this classification, they have not been included in Table \ref{t_a_SB1}. 

The parameters for the SB1 systems that may be synchronised are similar to those for the SB2 systems. For example, the $r$-estimates range from 0.19 to 0.72 with a median of 0.36. 
Additionally all the primaries appear to significantly fill their Roche lobe with \fl-estimates ranging from approximately 0.4 to 1.15 with a median of 0.75. Three systems have \fl-estimates that are greater than unity (\#189, 246, 662), but the uncertainties in the estimates of the stellar radii would by themselves be sufficient to allow \rl$<1$. 


A similar analysis can be undertaken for the sample of SB1 systems that appear to be asynchronous. This consists of all the remaining SB1 systems, excluding those with an upper limit to the projected rotational velocity as their synchronisation status remains uncertain. As the orbital inclination axis is unknown, we have assumed the mean $\sin i$-value for random inclination of axes of 0.785 \citep[see, for example,][]{gra05}. Now estimates for any individual primary will be uncertain but the analysis should allow the overall properties of the asynchronous systems to be inferred and the results are summarised in Table \ref{t_a_SB1_unsyn}.

\begin{table}
\caption{The estimated primary ($M_1$) and secondary masses ($M_2$),  semi-major axes ($a$) (units are \Msun\ and \Rsun), and fractional stellar radii ($r$) for the primaries in the SB1 sample that appear to be asynchronous. Also listed are their estimated primary Roche lobe radii (\rl\ in \Rsun) and the ratios to the stellar radii to their Roche lobe radii (\fl).}\label{t_a_SB1_unsyn}
{\tiny
\begin{center}
\begin{tabular}{llrrrrrr}
\hline\hline
VFTS & $P$(d) & $M_1$ & $M_2$ & $a$    & $r$  & \rl   & \fl  \\ \hline
015 &   8.79 & 13.6 &  2.9 &  45.6 & 0.138 &  22.5 & 0.28 \\
018 &  70.86 &  9.6 &  1.2 & 159.4 & 0.045 &  22.4 & 0.32 \\
037 &  41.24 & 10.0 &  1.1 & 112.0 & 0.081 &  49.0 & 0.19 \\
097 &  19.87 & 15.8 &  1.4 &  79.7 & 0.093 &  34.6 & 0.21 \\
106 &  16.35 & 13.4 &  1.3 &  66.5 & 0.083 &  20.1 & 0.28 \\
146 & 117.04 & 12.4 &  4.6 & 258.9 & 0.036 & 108.7 & 0.09 \\
155 & 153.43 & 11.8 &  4.4 & 304.6 & 0.015 & 110.8 & 0.04 \\
157 &  12.94 &  9.5 &  3.9 &  55.0 & 0.082 &  23.6 & 0.19 \\
162 & 145.43 &  9.9 &  3.8 & 277.8 & 0.023 &  69.5 & 0.09 \\
211 &  85.60 & 10.0 &  3.9 & 196.3 & 0.022 &  50.0 & 0.08 \\
213 &  13.59 & 11.1 &  1.2 &  55.3 & 0.182 &  11.4 & 0.88 \\
215 &   4.30 & 10.0 &  4.2 &  26.9 & 0.185 &   9.2 & 0.54 \\
218 &  20.75 & 15.4 &  1.2 &  81.1 & 0.196 &  21.7 & 0.73 \\
227 &  10.72 &  9.2 &  0.9 &  44.1 & 0.109 &  23.8 & 0.20 \\
257 & 132.25 & 10.2 &  3.4 & 260.3 & 0.017 &  94.8 & 0.05 \\
278 &  26.97 &  8.0 &  2.0 &  81.6 & 0.054 &  36.3 & 0.12 \\
334 &  38.20 & 11.0 &  2.7 & 114.2 & 0.047 &  49.8 & 0.11 \\
337 &  25.51 &  9.8 &  3.1 &  85.3 & 0.076 &  18.7 & 0.35 \\
359 &  19.49 & 13.1 &  2.4 &  76.0 & 0.078 &  20.9 & 0.28 \\
388 &  14.79 & 14.8 &  2.6 &  65.7 & 0.091 &  12.4 & 0.48 \\
396 &  86.03 & 13.2 &  3.0 & 207.1 & 0.032 &  55.5 & 0.12 \\
442 &  27.07 & 10.8 &  1.4 &  87.4 & 0.066 &  48.7 & 0.12 \\
501 & 151.46 & 10.8 &  2.1 & 280.0 & 0.033 & 135.1 & 0.07 \\
606 &  84.71 & 11.8 &  2.0 & 194.4 & 0.028 &  67.1 & 0.08 \\
665 &  23.54 & 13.3 &  3.4 &  88.2 & 0.072 &  19.1 & 0.33 \\
715 &   8.71 &  9.8 &  2.2 &  40.8 & 0.104 &  19.2 & 0.22 \\
718 &  26.54 &  8.2 &  2.8 &  83.1 & 0.085 &  28.9 & 0.24 \\
719 & 111.15 &  8.4 &  5.9 & 235.9 & 0.022 &  83.1 & 0.06 \\
723 &   9.95 & 14.2 &  2.3 &  49.5 & 0.157 &  26.7 & 0.29 \\
742 &   6.66 &  8.6 &  1.3 &  32.0 & 0.143 &  16.9 & 0.27 \\
784 &   3.58 & 12.7 &  0.8 &  23.4 & 0.264 &   6.5 & 0.95 \\
792 & 428.18 &  9.2 & 10.6 & 645.8 & 0.010 & 183.0 & 0.04 \\
821 &   9.50 & 12.6 &  1.8 &  45.9 & 0.107 &  24.5 & 0.20 \\
837 &   3.43 & 11.2 &  2.1 &  22.6 & 0.247 &  10.7 & 0.52 \\
874 & 370.82 & 11.2 &  4.4 & 543.1 & 0.019 & 197.9 & 0.05 \\
877 &  94.78 & 10.6 &  1.2 & 198.9 & 0.041 &  76.8 & 0.11 \\\hline
\end{tabular}
\end{center}
}
\end{table}

\begin{figure}
	\includegraphics[width=8.5cm, angle=0]{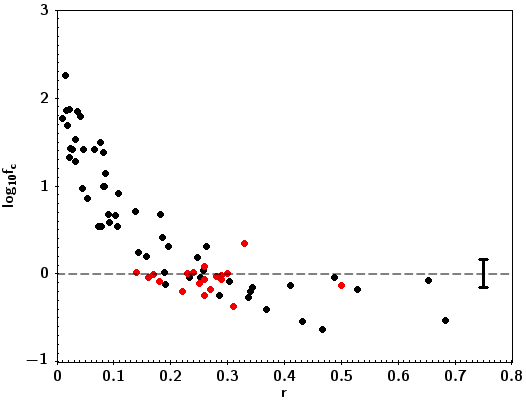} 
	
	\caption{Plots of \logfc-values for the primaries in the SB1 systems (black symbols) and primaries and secondaries in the SB2 systems (red symbols). Also shown is a typical \fci-error estimate and the upper limit for synchronous rotation assuming that the secondaries contribute negligible flux for the SB1 systems (dashed black line).}
	\label{f_r_a}
\end{figure} 

Fig.\ \ref{f_r_a} shows the variation of the \logfc-estimates against fractional radius, $r$. For the 36 \logfc-estimates for targets with $r > 0.2$, only five targets having \logfc-estimates greater than zero, with three of them lying within a 1$\sigma$\ uncertainty of this limit. The two remaining are SB1 systems and could have \logfc-estimates consistent with synchronisation by adopting a smaller $\sin i$-estimate. Hence we conclude that effectively all systems with $r > 0.2$ could be synchronised. As these system will generally have smaller periods, this is consistent with the results found in Sect.\ \ref{d_sep}.

Conversely all the 30 systems with $r < 0.15$ appear to be asynchronous even after including 1$\sigma$-uncertainties in the \logfc-estimates. For the 8 estimates with $0.15 < r < 0.2$, four are consistent with synchronisation, although all have relatively small \logfc-estimates implying that these systems might all have either reached, or be moving towards, synchronisation.

\subsection{Comparison with other observational and theoretical studies} \label{d_pre}

 Early studies \citep[see, for example,][]{pla59, lev74, lev76} of stellar rotation in binary systems found that synchronisation occurred up to a limiting orbital period estimated to be in the range of 4-10 days, consistent with the limit found in Sect.\ \ref{d_per}. \cite{lev74} did not find any correlation between the degree of synchronisation and the fractional radius, contrary to the results in Sect.\ \ref{d_sep}, whilst \citet{lev76} found that synchronisation extended to longer orbital periods with increasing stellar age. This was interpreted as due to tidal interactions during the main sequence stellar lifetimes. The binary systems in the VFTS sample will have a range of ages \citep[see, for example,][]{sch18} but will mostly be relatively young. Hence our relatively small limit of 3d for the orbital period of the synchronised systems would be compatible with the trend found by \cite{lev76}.
 
\citet{giu84ecl, giu84e, giu84l} have considered the degree of synchronisation in binary systems as a function of fractional radius\footnote{These authors do not define what is meant by their fractional radius and the current discussion assumes that it is the definition used here.}. \citet{giu84ecl} considered approximately 140 eclipsing systems that ranged in spectral type from effectively O7 to F6 (with one component classified as K-type). They found that almost all members were synchronised down to $r\sim0.15$ and both synchronised and asynchronous rotators were found down to $r \la 0.10$. There were insufficient systems to discuss the degree of synchronisation for $r < 0.1$. These results are comparable to those found here although we find a higher $r$-limit for synchronisation. However only 20 systems in \citet{giu84ecl} fell within our range of spectral types and as information was not provided for individual systems, no further comparison has been possible. 

Subsequently \citet{giu84e} analysed approximately 80 non-eclipsing SB2 systems with spectral types earlier than A5, of which 17 primaries had spectral types within our range of spectral types. They found that synchronisation occurred for fractional radii, $r\ga 0.05$, lower than that found in their previous analysis and significantly lower than has been found here. Given the relative small number of early B-type stars in their analysis and the absence of information for individual target, it is unclear where this discrepancy arises from.

The only recent studies of synchronisation have focused on lower mass stars. For instance \cite{zim17} and \cite{sai22} have studied synchronisation in terms of orbital period using r-mode oscillations in Kepler binary systems of later spectral types in the mass range 1.2 to 2.3 M$_\odot$.
Additionally, they generally have large eccentricities (ranging from $\sim$0.3--0.9), that lead to large corrections (up to a factor of 30) when estimating the periastron velocity, \vp. \citet{zim17} found that few of their sample of 24 systems were pseudo-synchronised. By contrast \citet{sai22} found pseudo-synchronisation for the majority of their twenty systems, many of which overlap with the sample of \cite{zim17}. However for two system the rotation periods were comparable to the corresponding orbital periods, whilst four systems did not appear to be synchronised. \citet{sai22} suggested that the differences in their conclusions compared with those of \citet{zim17}  might arise from different interpretations of the Fourier frequency spectra. 

Our results are consistent with those of \citet{sai22} for their shorter period systems although we find better agreement between the orbital and rotational periods than for the periastron values. Additionally \citet{sai22} find several system with orbital periods of more than 10 days that appear to be synchronised, contrary to our results. However given the lower masses and higher eccentricities of their sample compared to the one considered here, these differences are not surprising.

Few theoretical investigations of synchronisation in binary systems have been undertaken. \citet{zah77} considered the effects of tidal damping for both early- and late-type stars. For a system with equal mass components, lower limits for the initial fractional radius  and upper limits for the initial orbital period were estimated where synchronisation occurred within one quarter of the main sequence lifetime. For simulations with masses appropriate to early B-type stars these limits were $r\ga0.15$ and $P\la3.5$d. Subsequently \citet{hur02} found very similar limits for the initial fractional radius. These theoretical limits are in good agreement with those found here, namely that systems with periods greater than 10d or fractional radii less than 0.1 will be asynchronous, whilst those with periods of less than 3d and fractional radii greater than 0.2 are likely to be synchronised.

\begin{figure}
	\includegraphics[width=0.45\textwidth]{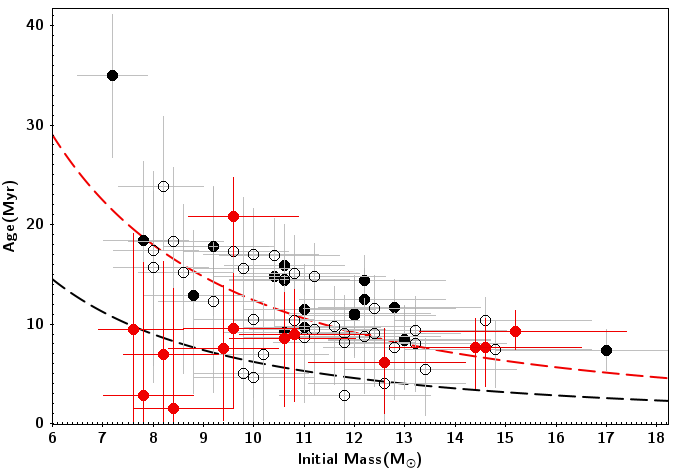}
	\caption{Current stellar age versus mass for the SB1 luminosity class III--V primaries (black circles, error bars in grey to aid legibility) and the SB2 primaries (red circles). Also shown are the loci of the 25\%\ (black dashed line) and 50\%\ (red dashed line) of main sequence lifetime ages as derived from the models of \cite{bro11a}. SB1 systems that are deemed synchronised (Table \ref{t_a_SB1}) are indicated as filled circles. All SB1 parameters were calculated assuming 0\,\%\ flux contribution of the secondary, however adopting a 20\,\%\ contribution only shifts points on average by approximately -0.5\,\Msun\ and +0.2\,Myr. }
	\label{f_a_m}
\end{figure}

Additionally, as {\sc bonnsai} produces estimates of the current age and initial mass for our SB1 luminosity class III--V sample we see, from Fig.\,\ref{f_a_m} (black circles), that our systems are older that the indicated 25\%\ of the main sequence lifetime required for synchronisation, as predicted by \cite{zah77}. 
Note that, of the three parameters used as input to {\sc bonnsai} the \vsini\ has little impact on the derived initial mass and current age, which are mainly determined by the current luminosity and effective temperature.
Therefore even for those systems that have synchronised this approach should yield reasonable results, provided the systems are not post-interaction binaries.
The same approach was used to estimate the ages of the SB2 primaries, with the same caveats, and these are also shown in Fig.\,\ref{f_a_m} (red circles) where one can see a comparable age distribution. 
There are two potential outliers for the SB2 systems that lie well below the 25\%\ boundary. However these are the circularised systems ($e=0$) \#199 and 752 that have positive error bars for their ages of 13.4 and 12.1 Myr respectively and hence their notional young ages are likely not significant.
The derived age distribution is also roughly consistent with their diffuse spatial distribution throughout the Tarantula Nebula as shown in Fig.1 of \cite{vil21}, and the star formation history of the overall region that indicates ongoing star formation over the past $\sim$20--30\,Myr \citep{sch18a,sab16}. For example very few, if any, of our systems belong to the central youngest cluster R\,136, that has an age of $\sim$2\,Myr. Consequently it is not surprising that our short period systems are synchronised. While this in good agreement with theory it implies that studies of younger binary systems are required in order to robustly test the timescale for synchronisation. The O-type binary systems in the Tarantula Nebula \citep{alm17,she22} may provide some additional insight for higher primary masses. 

\begin{acknowledgements}
Based on observations at the European Southern Observatory Very Large Telescope in programmes 171.D-0237, 182.D-0222 and 096.D-0825. DJL is supported by the Spanish Government Ministerio de Ciencia e Innovaci\'on through grants PID2021-122397NB-C21 and SEV 2015-0548. JIV acknowledges the European Research Council for support from the ERC Advanced grant ERC-2021-ADG101054731.
\end{acknowledgements}

%
%
\bibliographystyle{aa}
\bibliography{literature.bib}

\appendix
\section{SB2 systems}\label{a_sb2}
Below the individual characteristics of each of the SB2 systems are summarised but they share certain common features. In particular the parameters for the primary component are generally more secure than those of the secondary with in some cases no estimates being possible. Unless otherwise stated the \vsini-estimates were from the FT methodology.

\noindent\underline{\#112:} \vsini-estimates were possible for both components despite their relatively large values. The PF methodology was used in most cases although for the primary these were in good agreement with two FT estimates. Spectral types could be obtained for both components, which were consistent with the early B-type assignment of \citet{eva15}.

\noindent\underline{\#199:} \vsini-estimates and spectral types were obtained for both components. The latter were again consistent with the early B-type assignment of \citet{eva15}.

\noindent\underline{\#206:} \citet{vil21} defined the primary in SB2 systems as the star with the smaller K$_1$-estimate. For \#206, this appear to be the fainter star and as such here we have reversed the designations in Table \ref{t_f_sb2}.  \vsini-estimates could only be obtained for the primary and were near to the lower limit of 40\,\kms due to the spectral resolution. However estimates obtained from individual exposures were consistent with a small but finite rotational broadening. The spectral type of the primary was consistent with that of \citet{eva15}, although the luminosity class was uncertain due to possible blending with the secondary in the hydrogen and helium features.

\noindent\underline{\#240 and 883:} The spectroscopy for these target was relatively sparse with in both cases only 10 BBC exposures being used. Additionally the spectrum of \#240 contained emission lines. No parameters could be obtained for the secondaries whilst those for the primaries should be treated with caution. The spectral types of the primaries were consistent with that of \citet{eva15}.

\noindent\underline{\#248:} The relatively large flux ratio led to \vsini-estimates only being obtained for the primary. The strength of the \ion{O}{ii} spectrum led to a slightly earlier spectral type being assigned than that of \citet{eva15}

\noindent\underline{\#255:} \vsini-estimates and spectral types were obtained for both components.  For the primary, the latter agreed with that of \citet{eva15}.

\noindent\underline{\#364:} A \vsini-estimate was obtained for the primary but despite the low flux ratio, the secondary spectrum was not clearly identified. Broad features with central wavelengths consistent with a \ion{C}{ii} and \ion{He}{i} line may be present. If so the system could not be synchronised, which would be consistent with the relatively large \vsini-estimate for the primary. The spectral type of the primary agreed with that of \citet{eva15}.

\noindent\underline{\#383, 520 and 686:} The relatively large flux ratios led to a \vsini-estimates only being obtained for the primaries. The spectral types of the primaries agreed with those of \citet{eva15}

\noindent\underline{\#589:} Both components had very narrow lined metallic spectra leading to upper limits on their \vsini-values. The spectral type of the primary agreed with that of \citet{eva15}.

\noindent\underline{\#637:} The relatively large degree of rotational broadening made its estimation difficult and reliable values could only be estimated for the primary. The spectral type of the primary was consistent with that of \citet{eva15}.

\noindent\underline{\#752:} \vsini-estimates and spectral types were obtained for both components.  The strength of the \ion{O}{ii} spectrum led to a slightly earlier spectral type being assigned for the primary than that of \citet{eva15}

\section{SB1 data table}

\begin{table*}
\caption{Observed and derived parameters for the primaries in the SB1 sample, where \vc\ is the synchronous velocity as defined by equation\,1, with \logfc and \logfp being the inferred ratios of \vsini\ to \vc for the cases of circular and eccentric orbits respectively. The column headed `An' denotes the method used to derive the atmospheric parameters as discussed in subsection 3.2.}\label{t_SB1_full}
{\tiny
\begin{center}
\begin{tabular}{lccllrrrrrcrrcrr} 
\hline\hline
VFTS  &  V     &  (B-V) &  Sp.     &  Class    & \vsini & \teff  & An & E(B-V) & BC    &  log\,$L/L_{\odot}$   &  $P$(d)   &  $R(R_{\odot})$   &  \vc   &  \logfc  & \logfp\\
 & & & Type & & \kms & 1000\,K & & & & & & & \kms & &  \\ \hline
009 & 16.18 & 0.11 & B1-1.5 V & SB1 & 87 & 26200 & ST & 0.36 & -2.60 & 4.37 & 4.71 & 7.42 & 79.70 & 0.04 & 0.03 \\
015 & 16.2 & 0.04 & B0.5 V & SB1 & 185 & 28000 & He & 0.30 & -2.76 & 4.34 & 8.79 & 6.32 & 36.37 & 0.71 & 0.70 \\
018 & 16.62 & 0.17 & B1.5 V & SB1* & 48 & 23000 & G & 0.39 & -2.28 & 4.11 & 70.86 & 7.17 & 5.12 & 0.97 & 0.44 \\
027 & 14.71 & 0.09 & B1 III-II & SB1 & 91 & 20500 & M & 0.29 & -2.00 & 4.61 & 6.58 & 16.14 & 124.06 & -0.13 & -0.13 \\
033 & 16.2 & 0.07 & B1-1.5 V & SB1 & 77 & 24000 & G & 0.30 & -2.38 & 4.19 & 3.86 & 7.24 & 95.00 & -0.09 & -0.09 \\
037 & 15.79 & 0.06 & B2 III: & SB1* & 267 & 21700 & ST & 0.27 & -2.14 & 4.21 & 41.24 & 9.08 & 11.14 & 1.38 & 1.26 \\
041 & 16.94 & 0.14 & B2: V & SB1 & -40 & 23000 & G & 0.36 & -2.28 & 3.94 & 14.33 & 5.90 & 20.82 & 0.28 & 0.10 \\
097 & 16.21 & 0.16 & B0 IV & SB1 & 72 & 29500 & He & 0.43 & -2.89 & 4.57 & 19.87 & 7.38 & 18.81 & 0.58 & 0.43 \\
106 & 16.43 & 0.05 & B0.2 V & SB1* & 170 & 30000 & He & 0.32 & -2.93 & 4.35 & 16.35 & 5.53 & 17.11 & 1.00 & 0.58 \\
146 & 16.24 & 0.25 & B2: V & SB1* & 287 & 24500 & ST & 0.49 & -2.43 & 4.45 & 117.04 & 9.39 & 4.06 & 1.85 & 1.82 \\
155 & 16.95 & 0.08 & B0.7: V & SB1* & 277 & 29000 & He & 0.35 & -2.85 & 4.14 & 153.43 & 4.67 & 1.54 & 2.26 & 2.14 \\
157 & 16.69 & -0.08 & B2 V & SB1 & 175 & 24500 & ST & 0.16 & -2.43 & 3.81 & 12.94 & 4.48 & 17.54 & 1.00 & 0.99 \\
162 & 16.49 & 0.07 & B0.7 V & SB1 & 60 & 23000 & G & 0.29 & -2.28 & 4.02 & 145.43 & 6.48 & 2.25 & 1.43 & 1.03 \\
179 & 16.93 & 0.09 & B1 V & SB1 & 51 & 27000 & G & 0.34 & -2.67 & 4.08 & 1.16 & 5.00 & 217.55 & -0.63 & -0.66 \\
189 & 16.31 & 0.1 & B0.7: V & SB1 & 212 & 27900 & ST & 0.36 & -2.75 & 4.38 & 1.43 & 6.63 & 233.77 & -0.04 & -0.05 \\
195 & 16.86 & 0.06 & B0.5 V & SB1 & -40 & 28000 & G & 0.32 & -2.76 & 4.11 & 15.01 & 4.81 & 16.23 & 0.39 & 0.36 \\
204 & 16.13 & 0.31 & B2 III & SB1 & -40 & 22500 & G & 0.53 & -2.23 & 4.47 & 164.34 & 11.39 & 3.51 & 1.06 & 0.42 \\
211 & 16.56 & -0.12 & B1 V & SB1 & 188 & 26800 & ST & 0.13 & -2.65 & 3.92 & 85.60 & 4.24 & 2.51 & 1.87 & 1.50 \\
213 & 15.5 & 0.04 & B2 III:e & SB1* & 181 & 21700 & ST & 0.25 & -2.14 & 4.30 & 13.59 & 10.04 & 37.40 & 0.68 & 0.00 \\
215 & 16.58 & -0.03 & B1.5 V & SB1 & 154 & 25700 & ST & 0.22 & -2.55 & 3.99 & 4.30 & 4.98 & 58.55 & 0.42 & 0.28 \\
218 & 15.63 & 0.38 & B1.5 V & SB1* & 79 & 23000 & G & 0.60 & -2.28 & 4.80 & 20.75 & 15.87 & 38.70 & 0.31 & -0.22 \\
225 & 15.07 & -0.01 & B0.7-1 III-II & SB1 & 59 & 24500 & G & 0.23 & -2.43 & 4.56 & 8.24 & 10.58 & 65.05 & -0.04 & -0.04 \\
227 & 16.51 & -0.09 & B2 V & SB1 & 183 & 24500 & ST & 0.15 & -2.43 & 3.87 & 10.72 & 4.79 & 22.63 & 0.91 & 0.89 \\
246 & 16.83 & 0.45 & B1 III & SB1 & 135 & 24200 & ST & 0.68 & -2.40 & 4.48 & 2.44 & 9.94 & 206.06 & -0.18 & -0.18 \\
257 & 16.7 & -0.04 & B0.7-1.5V & SB1 & 125 & 26800 & ST & 0.21 & -2.65 & 3.98 & 132.25 & 4.53 & 1.73 & 1.86 & 1.73 \\
278 & 16.82 & -0.07 & B2.5 V & SB1 & 60 & 23400 & ST & 0.16 & -2.32 & 3.72 & 26.97 & 4.40 & 8.26 & 0.86 & 0.83 \\
291 & 14.85 & 0.12 & B5 II-Ib & SB1 & -20 & 13500 & M & 0.23 & -0.98 & 4.07 & 107.46 & 19.79 & 9.32 & 0.33 & 0.33 \\
299 & 16.36 & -0.05 & B0.5 V & SB1 & -40 & 28000 & G & 0.21 & -2.76 & 4.15 & 20.26 & 5.07 & 12.68 & 0.50 & 0.05 \\
305 & 16.59 & -0.1 & B2: V & SB1 & 57 & 24500 & ST & 0.14 & -2.43 & 3.82 & 4.18 & 4.55 & 55.06 & 0.02 & 0.01 \\
324 & 15.53 & -0.13 & B0.2 V & SB1 & 57 & 28500 & G & 0.13 & -2.80 & 4.39 & 1.64 & 6.47 & 199.24 & -0.54 & -0.55 \\
334 & 16.26 & -0.06 & B0.7 V & SB1 & 182 & 27000 & He & 0.19 & -2.67 & 4.13 & 38.20 & 5.34 & 7.08 & 1.41 & 1.37 \\
337 & 16.72 & 0.14 & B2: V-IIIe+ & SB1* & 412 & 23000 & ST & 0.36 & -2.28 & 4.03 & 25.51 & 6.52 & 12.95 & 1.50 & 0.98 \\
342 & 16.94 & -0.04 & B1 V & SB1 & -40 & 23000 & G & 0.18 & -2.28 & 3.69 & 4.28 & 4.41 & 52.19 & -0.12 & -0.14 \\
351 & 15.98 & 0.08 & B0.5 V & SB1 & -40 & 28500 & G & 0.34 & -2.80 & 4.51 & 67.40 & 7.37 & 5.54 & 0.86 & 0.74 \\
359 & 16.3 & 0.03 & B0.5 V & SB1 & 54 & 28000 & G & 0.29 & -2.76 & 4.29 & 19.49 & 5.93 & 15.41 & 0.54 & 0.13 \\
388 & 16.42 & 0.06 & B0.5 V & SB1* & 98 & 27500 & He & 0.32 & -2.72 & 4.26 & 14.79 & 5.96 & 20.38 & 0.68 & -0.01 \\
396 & 16.32 & 0.11 & B0.5 V & SB1 & 131 & 28500 & He & 0.37 & -2.80 & 4.41 & 86.03 & 6.62 & 3.89 & 1.53 & 1.11 \\
430 & 15.11 & 0.64 & B0.5 Ia+ ((n))Nwk & SB1 & 98 & 24500 & M & 0.88 & -2.43 & 5.45 & 8.76 & 29.63 & 171.17 & -0.24 & -0.24 \\
434 & 16.13 & 0.16 & B1.5: V & SB1 & 45 & 23000 & G & 0.38 & -2.28 & 4.29 & 5.61 & 8.84 & 79.83 & -0.25 & -0.26 \\
442 & 16.66 & 0.08 & B1-2 V & SB1* & 281 & 25700 & ST & 0.33 & -2.55 & 4.11 & 27.07 & 5.73 & 10.71 & 1.42 & 1.42 \\
501 & 15.74 & 0.08 & B0.5 V & SB1 & 59 & 23000 & G & 0.30 & -2.28 & 4.34 & 151.46 & 9.30 & 3.11 & 1.28 & 1.26 \\
534 & 15.66 & 0.21 & B0 IV & SB1 & 57 & 29000 & G & 0.48 & -2.85 & 4.84 & 3.69 & 10.42 & 143.15 & -0.40 & -0.41 \\
575 & 15.11 & 0.02 & B0.7 III & SB1 & -40 & 26000 & G & 0.27 & -2.58 & 4.66 & 45.30 & 10.55 & 11.79 & 0.53 & 0.10 \\
576 & 15.67 & 0.78 & B1 IaNwk & SB1* & 52 & 20000 & M & 0.97 & -1.94 & 5.17 & 85.60 & 31.94 & 18.89 & 0.44 & -0.14 \\
606 & 16.60 & 0.06 & B0-0.5 V(n) & SB1* & 86 & 28000 & He & 0.32 & -2.76 & 4.21 & 84.71 & 5.42 & 3.24 & 1.42 & 1.17 \\
662 & 16.12 & 0.08 & B3-5 III: & SB1* & 67 & 17500 & G & 0.24 & -1.61 & 3.83 & 1.99 & 9.00 & 229.03 & -0.53 & -0.53 \\
665 & 16.43 & 0.11 & B0.5 V & SB1 & 47 & 28000 & G & 0.37 & -2.76 & 4.35 & 23.54 & 6.36 & 13.67 & 0.54 & -0.02 \\
687 & 14.29 & 0.28 & B1.5 Ib((n))Nwk & SB1 & 123 & 20000 & M & 0.47 & -1.94 & 5.02 & 12.47 & 26.94 & 109.35 & 0.05 & 0.05 \\
705 & 16.43 & 0.07 & B0.7 V & SB1 & 87 & 26000 & He & 0.32 & -2.58 & 4.20 & 2.26 & 6.23 & 139.41 & -0.20 & -0.21 \\
715 & 16.64 & -0.1 & B1 V & SB1 & 116 & 26800 & ST & 0.15 & -2.65 & 3.92 & 8.71 & 4.22 & 24.54 & 0.67 & 0.66 \\
718 & 15.99 & -0.06 & B2.5 III & SB1 & 185 & 20400 & ST & 0.14 & -1.99 & 3.89 & 26.54 & 7.05 & 13.44 & 1.14 & 0.98 \\
719 & 17.00 & 0.08 & B1 V & SB1 & 50 & 23000 & G & 0.30 & -2.28 & 3.83 & 111.15 & 5.21 & 2.37 & 1.32 & 1.28 \\
723 & 16.19 & 0.16 & B0.5 V & SB1 & 63 & 27500 & G & 0.42 & -2.72 & 4.49 & 9.95 & 7.78 & 39.57 & 0.20 & 0.20 \\
730 & 15.41 & -0.1 & B1 IV(n) & SB1* & 248 & 25500 & ST & 0.14 & -2.53 & 4.35 & 1.33 & 7.66 & 290.43 & -0.07 & -0.07 \\
742 & 16.93 & -0.02 & B2 V & SB1 & 60 & 23000 & G & 0.20 & -2.28 & 3.72 & 6.66 & 4.58 & 34.76 & 0.24 & 0.23 \\
779 & 15.46 & 0.19 & B1 II-Ib & SB1 & 47 & 23500 & M & 0.42 & -2.33 & 4.63 & 59.93 & 12.49 & 10.55 & 0.65 & 0.64 \\
784 & 16.83 & 0.19 & B1: V & SB1* & 180 & 26800 & ST & 0.44 & -2.65 & 4.25 & 3.58 & 6.18 & 87.41 & 0.31 & -0.21 \\
788 & 16.15 & 0.09 & B1 III & SB1 & 83 & 24100 & ST & 0.32 & -2.39 & 4.25 & 3.26 & 7.63 & 118.43 & -0.15 & -0.15 \\
792 & 15.96 & -0.06 & B2 V & SB1 & 47 & 23000 & G & 0.16 & -2.28 & 4.05 & 428.18 & 6.71 & 0.79 & 1.77 & 1.65 \\
799 & 16.86 & 0.1 & B0.5-0.7 V & SB1* & -40 & 26500 & G & 0.35 & -2.63 & 4.09 & 39.37 & 5.30 & 6.82 & 0.77 & -0.02 \\
821 & 16.03 & -0.14 & B0 V-IV & SB1 & 91 & 30000 & He & 0.13 & -2.93 & 4.24 & 9.50 & 4.89 & 26.09 & 0.54 & 0.54 \\
827 & 15.34 & 0.31 & B1.5 Ib & SB1 & 52 & 21000 & M & 0.51 & -2.06 & 4.70 & 43.229 & 17.00 & 19.90 & 0.42 & 0.26 \\
829 & 15.13 & 0.41 & B1.5-2 II & SB1 & 138 & 20500 & M & 0.61 & -2.00 & 4.89 & 202.93 & 22.27 & 5.56 & 1.40 & 1.23 \\
837 & 16.07 & -0.09 & B1 V & SB1 & 129 & 26800 & ST & 0.16 & -2.65 & 4.16 & 3.43 & 5.58 & 82.37 & 0.19 & 0.17 \\
850 & 16.15 & 0.18 & B1 III & SB1* & -40 & 24000 & G & 0.41 & -2.38 & 4.37 & 50.66 & 8.84 & 8.83 & 0.66 & 0.61 \\
874 & 15.37 & 0.02 & B1.5 IIIe+ & SB1 & 67 & 22950 & ST & 0.24 & -2.27 & 4.40 & 370.82 & 10.02 & 1.37 & 1.69 & 1.59 \\
877 & 16.36 & 0.18 & B1-3 V-IIIe+ & SB1* & 267 & 23125 & ST & 0.40 & -2.29 & 4.24 & 94.78 & 8.19 & 4.37 & 1.79 & 1.57 \\
888 & 16.18 & -0.07 & B0.5 V & SB1 & 76 & 27000 & G & 0.18 & -2.67 & 4.15 & 1.97 & 5.45 & 140.44 & -0.27 & -0.27 \\
891 & 16.48 & 0.07 & B2 V & SB1 & 55 & 23000 & G & 0.29 & -2.28 & 4.03 & 5.48 & 6.51 & 60.11 & -0.04 & -0.20 \\
\hline
\end{tabular}
\end{center}
}
\end{table*}

\end{document}